\documentclass[iop, revtex4]{emulateapj}
\usepackage[breaklinks,colorlinks,citecolor=blue,linkcolor=blue]{hyperref}
\usepackage{amsmath}

\usepackage{etoolbox}

\makeatletter

\patchcmd{\NAT@citex}
  {\@citea\NAT@hyper@{%
     \NAT@nmfmt{\NAT@nm}%
     \hyper@natlinkbreak{\NAT@aysep\NAT@spacechar}{\@citeb\@extra@b@citeb}%
     \NAT@date}}
  {\@citea\NAT@nmfmt{\NAT@nm}%
   \NAT@aysep\NAT@spacechar\NAT@hyper@{\NAT@date}}{}{}

\patchcmd{\NAT@citex}
  {\@citea\NAT@hyper@{%
     \NAT@nmfmt{\NAT@nm}%
     \hyper@natlinkbreak{\NAT@spacechar\NAT@@open\if*#1*\else#1\NAT@spacechar\fi}%
       {\@citeb\@extra@b@citeb}%
     \NAT@date}}
  {\@citea\NAT@nmfmt{\NAT@nm}%
   \NAT@spacechar\NAT@@open\if*#1*\else#1\NAT@spacechar\fi\NAT@hyper@{\NAT@date}}
  {}{}

\makeatother

\newcommand{\myemail}{tleung@astro.cornell.edu}
\newcommand{\Msun}{\mbox{$M_{\odot}$}}

\newcommand{\Lsun}{\mbox{L$_{\odot}$}}
\newcommand{\rarr}{$\rightarrow$}
\newcommand{\CO}{\mbox{CO($J$\,=\,3\,$\rightarrow$\,2) }}
\newcommand{\rot}[3][CO]{\mbox{#1($J$\,=\,#2\,\rarr\,#3)}}
\newcommand{\Lp}{\mbox{$L^{\prime}_\textrm{CO(1-0)}$}}
\newcommand{\LpU}{\mbox{K\,\,km\,\,s$^{-1}$\,\,pc$^2$}}
\newcommand{\eg}{{\sl e.g.,~}}
\newcommand{\ie}{{\sl i.e.,~}}
\newcommand{\pmOne}{\mbox{$^{-1}$}}
\newcommand\tna{\,\tablenotemark{a}}
\newcommand\tnb{\,\tablenotemark{b}}
\newcommand\tnc{\,\tablenotemark{c}}
\newcommand\tnd{\,\tablenotemark{d}}
\newcommand\tne{\,\tablenotemark{e}}
\newcommand\tnf{\,\tablenotemark{f}}
\newcommand\tng{\,\tablenotemark{g}}
\newcommand\tnh{\,\tablenotemark{h}}
\newcommand\tni{\,\tablenotemark{i}}
\newcommand\tnj{\,\tablenotemark{j}}
\newcommand\tnk{\,\tablenotemark{k}}
\newcommand\tnl{\,\tablenotemark{l}}

\slugcomment{Accepted to the ApJ}

\shorttitle{Study of a strongly-lensed type-2 quasar host SMG at $z$\,=\,2.221}
\shortauthors{Leung \& Riechers}

\begin{document}
\title{A Massive Molecular Gas Reservoir in the $z$\,=\,2.221 Type-2 Quasar Host Galaxy SMM\,J0939+8315 lensed by the Radio Galaxy 3C220.3}
\author{T. K. Daisy Leung and Dominik A. Riechers}
\affil{Department of Astronomy, Space Sciences Building, Cornell University, Ithaca, NY 14853, USA; \myemail}

\begin{abstract}
We report the detection of \CO line emission in the strongly-lensed submillimeter galaxy (SMG) SMM\,J0939+8315 at $z$ = 2.221, using
the Combined Array for Research in Millimeter-wave Astronomy.
SMM\,J0939+8315 hosts a type-2 quasar, and is gravitationally lensed by the radio galaxy 3C220.3 and its companion galaxy at $z$ = 0.685.
The 104~GHz continuum emission underlying the CO line is detected toward 3C220.3 with an integrated flux density of $S_\textrm{cont}$ = 7.4\,$\pm$\,1.4 mJy.
Using the \CO line intensity of $I_\textrm{CO(3-2)}$ = (12.6\,$\pm$\,2.0)\,Jy\,km\,s\pmOne, we derive
 a lensing- and excitation-corrected CO line luminosity of \Lp = (3.4\,$\pm$\,0.7)\,$\times$\,10$^{10}$\,(10.1/$\mu_\textrm{L}$)\,\LpU\ for the SMG, where $\mu_\textrm{L}$ is the lensing magnification factor inferred from our lens modeling.
 This translates to a molecular gas mass of $M_\textrm{gas}$ = (2.7\,$\pm$\,0.6)\,$\times$\,10$^{10}$\,(10.1/$\mu_\textrm{L}$)\,\Msun.
Fitting spectral energy distribution models to the (sub)-millimeter data of this SMG yields a dust temperature of $T$\,=\,63.1$^{+1.1}_{-1.3}$\,K, a dust mass of $M_\textrm{dust}$ = (5.2\,$\pm$\,2.1)\,$\times$\,10$^8$\,(10.1/$\mu_\textrm{L}$)\,\Msun, and a total infrared luminosity of
$L_\textrm{IR}$ = (9.1\,$\pm$\,1.2)\,$\times$10$^{12}$\,(10.1/$\mu_\textrm{L}$)\,\Lsun.
We find that the properties of the interstellar medium of
 SMM J0939+8315 overlap with both SMGs and type-2 quasars. Hence, SMM J0939+8315 may be transitioning from a star-bursting phase to an unobscured quasar phase as described by the
``evolutionary link'' model, according to which this system may represent an intermediate stage in the evolution of present-day galaxies
at an earlier epoch.
\end{abstract}
\keywords{cosmology: observations --- galaxies: evolution --- galaxies: high-redshift ---  galaxies: starburst --- submillimeter: galaxies}

\section{Introduction}\label{sec:intro}
Submillimeter-selected galaxies (SMGs) are predominantly found at redshifts $z$\,$\sim$\,1\,--\,3 \citep{Chapman05a}, during the epoch of stellar mass and
galaxy assembly, with a tail out to $z>$ 6 \citep{Riechers13a}.
Previous works
have shown that SMGs
 are extremely luminous in the infrared
 wavelengths (L$_\textrm{IR} \sim$ 10$^{12}$ \Lsun) with high star formation rates \citep[SFR $\gtrsim$\,500\,\Msun yr\pmOne; see \eg reviews by][]{Blain02a, Lagache05a,Casey14a}.
Following the pioneering works in the discovery of this population \citep{Smail97a,Hughes98a,Barger98a}, considerable amounts of effort have been invested into
obtaining large samples of SMGs by carrying out
  large sky surveys with (sub)-mm facilities such as the {\it Herschel Space Observatory} \citep[\eg H-ATLAS, SPT, HerMES; ][]{Eales10a,Carlstrom11a,Oliver12a}.
 
  To characterize the physical properties of the gas reservoirs in the interstellar medium (ISM) where active star formation takes place, carbon monoxide ($^{12}$CO) rotational lines have been commonly used as tracers due to its high abundance in the ISM as well as its low excitation energy; the ground state transition line thereby directly probes the cool gas that is essential to fuel star formation \citep[see \eg reviews by][]{Solomon05a,Carilli13a}. Observations of CO in SMGs
  have demonstrated that these galaxies have large gas reservoirs typical of \textgreater 10$^{10}$\Msun \citep[\eg][]{Frayer98a, Neri03a, Riechers11c, Riechers11d, Ivison11a, Bothwell13a}.

Many recent detailed studies have been carried out on SMGs that are gravitationally lensed,
 as lensing amplifies the intrinsic fluxes of these sources, making them the brightest unveiled in large sky surveys \citep{Negrello10a,Vieira10a,Oliver12a}, and making follow-up studies considerably less time consuming.
A particularly interesting and peculiar lensing system was discovered serendipitously in a study carried out with the {\it Herschel Space Observatory}, in which
a type-2 quasar host SMG --- SMM\,J0939+8315 (hereafter SMM\,J0939) is being lensed by the double-lobed Fanaroff-Riley
Class II \citep*[FR-II; ][]{Fanaroff74} radio galaxy 3C220.3 at $z$\,=\,0.685, which has a
companion galaxy ``B'' as detected in Keck 2.2$\micron$ and the {\it Hubble Space Telescope} 702nm images \citep[hereafter H14]{Haas14}.
SMM\,J0939 is currently one of the brightest known lensed
SMGs, with a lensing-magnified flux density of $S_\textrm{250\micron}$\,=\,440\,$\pm$\,15 mJy.
Detections of C~{\scriptsize\sc IV}~1549\AA\
 and He~{\scriptsize\sc II}~1640\AA\ line emission toward SMM\,J0939
 place the redshift of this galaxy at $z$\,=\,2.221. Based on the spectral line fluxes and line widths, H14 suggest the presence of an obscured active galactic nucleus (AGN) in the form of a type-2 quasar in this SMG.

In this paper, we present the detection of \CO line emission toward the background SMG obtained with the Combined
Array for Research in Millimeter Astronomy (CARMA), which confirms and refines the redshift, and permits a study of the physical conditions in the ISM of SMM\,J0939 in great detail. We report the detection of the continuum emission underlying the CO line and place constraints on the spectral energy distribution (SED) of the foreground FR-II galaxy at millimeter (mm) wavelengths ($\sim$\,104 GHz). Based on the magnification factor derived from our lens model, we infer various intrinsic properties of SMM\,J0939. We
conclude this paper by comparing our findings to other similarly bright, strongly-lensed SMGs, as well as other type-2 quasars at $z$\,$\sim$\,2$-$3.

We adopt a flat $\Lambda$CDM cosmological model throughout this paper, with H$_0$= 69.32 km\,\,Mpc\pmOne\,\,s\pmOne, $\Omega_\textrm{M}$\,=\,0.286, $\Omega_\Lambda$=0.713, based on the WMAP9 results \citep{Hinshaw13a}.
The luminosity distances at $z$\,=\,0.685 and $z$\,=\,2.221 are 4214 Mpc and 18052 Mpc, respectively; 1$\arcsec$
corresponds to 7.169 kpc at $z$\,=\,0.685, and 8.406 kpc at $z$\,=\,2.221.

\section{Observations}\label{sec:obs}
\subsection{CARMA} \label{sec:carmadata}
Observations of the \CO rotational transition ($\nu_\textrm{rest}$\,=\,345.7959899 GHz) toward the background galaxy SMM
J0939 ($z$\,=\,2.221) were carried out using CARMA at a redshifted frequency of $\nu_{\rm obs}$\,=\,107.357\,\,GHz (2.79\,\,mm; program ID: cf0142; PI: Riechers). The 3\,mm receivers were used to cover the redshifted \CO line and the nearby observed-frame 2.88\,mm continuum emission. The correlator was configured to provide an effective bandwidth of 3.708 GHz in each sideband, and a spectral resolution of 5.208 MHz ($\sim$\,14.5 km\,\,s\pmOne).
The line was placed in the
upper sideband, with the local oscillator tuned to $\nu_\textrm{LO}$\,$\sim$\,104.2609 GHz.
Observations were carried out under good
weather conditions in the E array configuration on 2014 July 12. This resulted in 1.56 hours of 15 antenna-equivalent on-source time after discarding unusable visibility data.
The nearby source J1039+811 (0.65\,\,Jy) was observed every 20 minutes for
pointing, amplitude, and phase calibration. Mars was observed as the primary
absolute flux calibrator, and the quasar 3C273 was observed as the secondary
flux calibrator. J0927+390 was observed for bandpass calibration, yielding $\sim
$15\% calibration accuracy. \par
We use the {\sc miriad} package to calibrate and analyze the visibility data, which are deconvolved and imaged with ``natural" weighting.
This yields a synthesized clean beam size of 11$\farcs$5\,$\times$\,6\farcs2, $-$56.1$\degr$ east of north for the upper sideband image cube and an rms noise of $\sigma_\textrm{ch}$\,=\,9.49\,\,mJy\,\,beam\pmOne\ per channel
of width $\sim$\,29 km\,\,s\pmOne.
The continuum image is created by
averaging over all line-free channels; this yields a synthesized clean beam of 12\farcs0\,$\times$\,6\farcs5, $-$55.9$\degr$ east of north, and an rms noise of $\sigma_\textrm{cont}$ = 0.50\,\,mJy\,\,beam\pmOne\ over 6.8 GHz.

\section{Results}\label{sec:res}
\subsection{Foreground Galaxy: 3C220.3}
\begin{figure*}[tbph]
\centering
\includegraphics[width=0.80\textwidth]{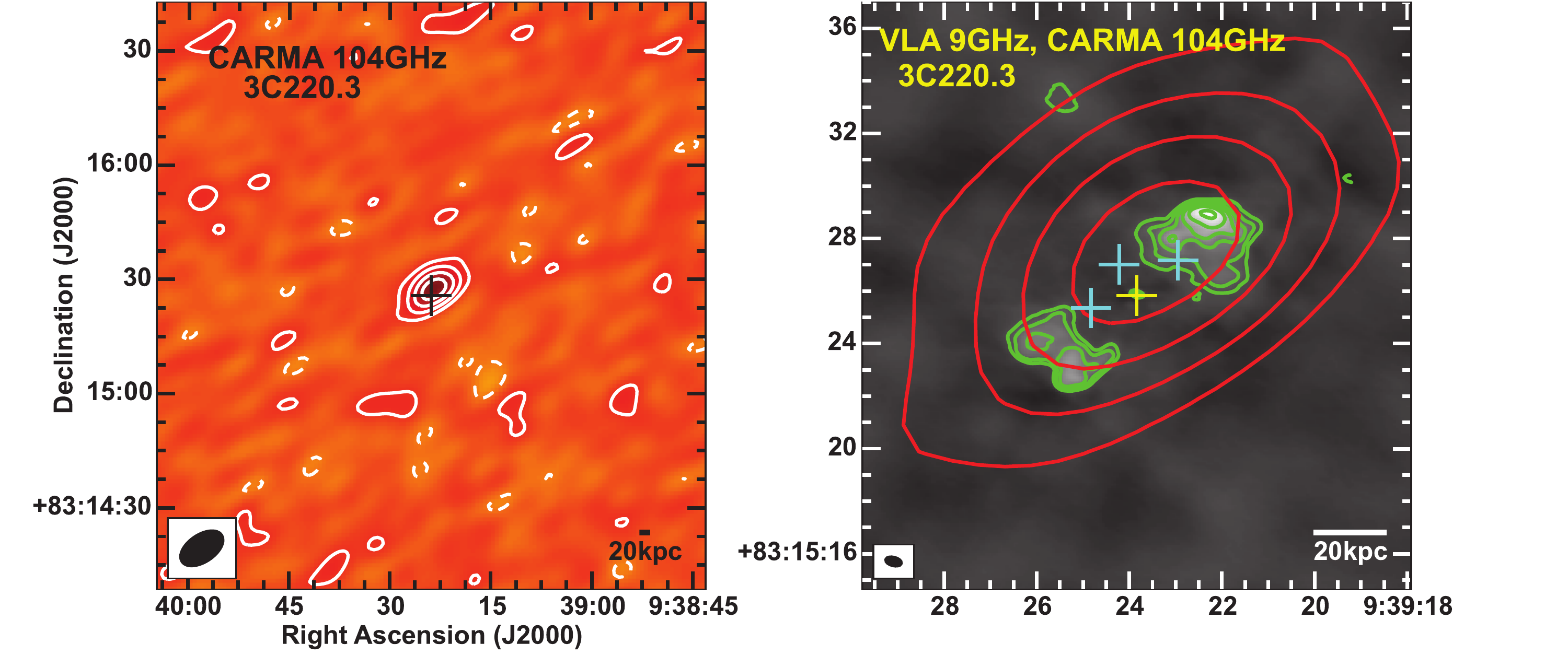}
\caption{Left: Contour map of the 104 GHz continuum emission in the foreground radio galaxy 3C220.3.
The beam size is 12\farcs0\,$\times$\,6\farcs5, at P.A.\,=\,
$-$56$\degr$, as indicated in the bottom left corner. Right: CARMA 104 GHz continuum emission (red contours) overlaid on the VLA 9 GHz continuum emission (green contours and grayscale; H14).
The synthesized beam size of the VLA observations is 0$\farcs$6\,$\times$\,0$\farcs$2, at P.A.
76$\degr$. The contour levels of the 104 GHz continuum emission start at $\pm$2$\sigma$, incrementing at steps
of $\pm$2$\sigma$, where $\sigma$ = 0.5 mJy beam\pmOne. The contour levels of the 9 GHz continuum
emission start at $\pm$4$\sigma$, where $\sigma$\,=\,0.064 mJy beam\pmOne, and increment at steps of $\pm$2$^n\sigma$,
where $n$ is a positive integer. The blue crosses correspond to the centroid locations of the lensing knots detected in the SMA 1\,mm continuum emission (see Figure \ref{fig:mom0}). The central cross on each panel indicates the position of the radio core of 3C220.3. \label{fig:cont}}
\end{figure*}

Averaging over all line-free channels, we detect continuum emission at $\sim$\,9$\sigma$ significance at an averaged frequency of $\nu_\textrm{cont}$\,=\,104.2106 GHz ($\sim$\,2.9 mm) in the observed-frame, corresponding to 175.6 GHz ($\sim$\,1.7 mm) at $z$\,=\,0.685. In this lensing system, the
foreground galaxy (3C220.3) is radio-loud, we thus expect it to be the dominant contributor to the continuum emission (see \S \ref{sec:SEDFg} for details). The task {\sc imfit} is used to estimate the peak position of the continuum emission, where the flux density is S$_\nu$\,=\,4.93\,$\pm$\,0.31\,\,mJy\,\,beam\pmOne. From the continuum measurement, the deconvolved source size
is (8\farcs4\,$\pm$\,1\farcs1)\,$\times$\,(4\farcs9\,$\pm$\,0\farcs6) at $-$53.8$\degr$, and the integrated flux density is 7.39\,$\pm$\,1.42\,\,mJy. An overlay image of the 104 GHz
continuum emission with the 9 GHz continuum emission (H14) is shown in Figure~\ref{fig:cont}, demonstrating that the continuum
emission is marginally resolved at the resolution of our observations. It is therefore plausible that non-thermal emission from the radio lobes and core of the foreground galaxy dominate the integrated flux
density of the measured continuum. We discuss this further in Section \ref{sec:SEDFg}. \par
The frequency range of our observations covers the HCO$^+$($J$\,=\,2\,\rarr\,1), HNC($J$\,=\,2\,$\rightarrow$\,1), and H$_2$O(3$_{13}$\,\rarr\,2$_{20}$)
transition lines in the foreground galaxy, at
the redshifted frequencies of 105.86, 107.71, and 108.79\,\,GHz, respectively. We establish 3$\sigma$ upper limits employing a typical FWHM line width of
$\sim$\,300\,\,km\,\,s\pmOne, based on the CO($J$\,=\,1\,$\rightarrow$\,0) line measurements in a sample of local radio galaxies \citep[$z$ $<$ 0.1; ][]{Smolcic11a}. This results in upper limits of $<$ 2.66\,\,Jy\,\,km\,\,s\pmOne\ on the integrated emission line strengths.

\subsection{Background Galaxy: SMM\,J0939}
\begin{figure*}[tbph]
\centering
\includegraphics[width=0.8\textwidth]{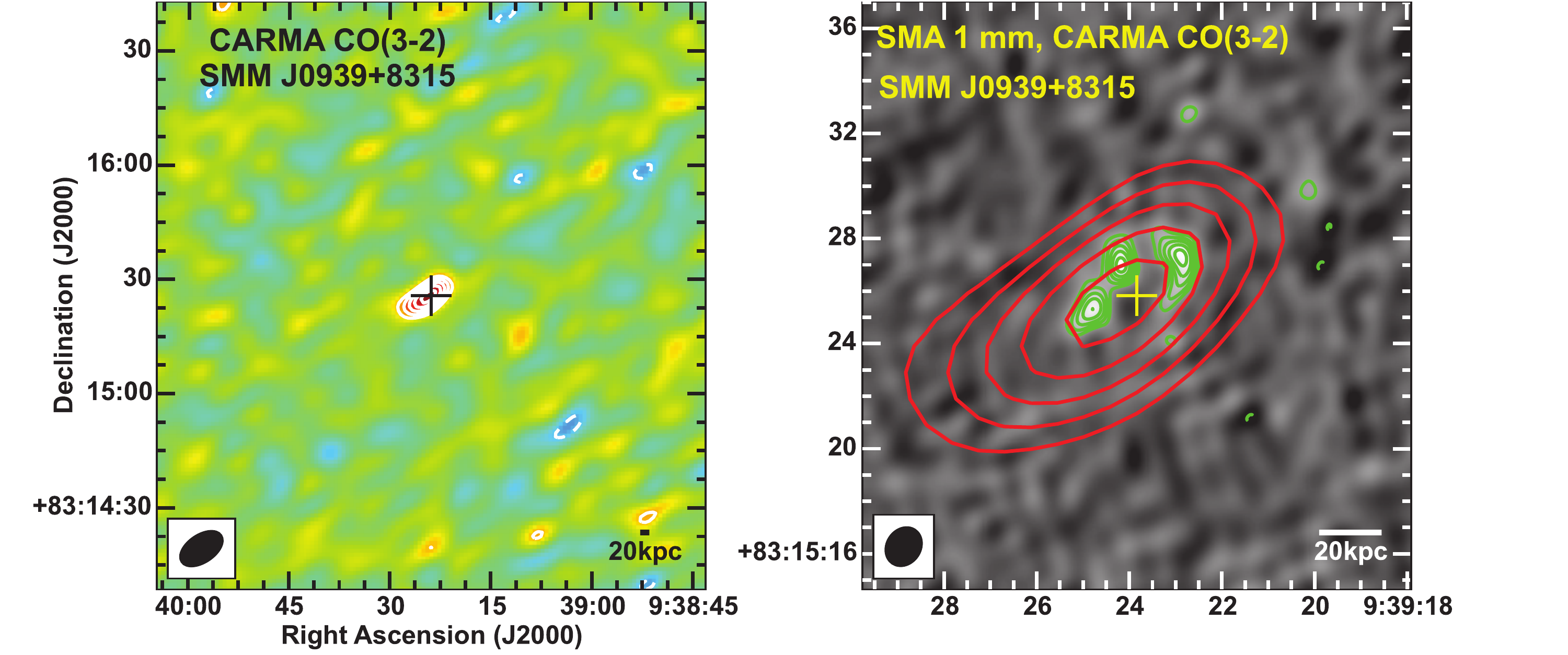}
\includegraphics[width=0.65\textwidth]{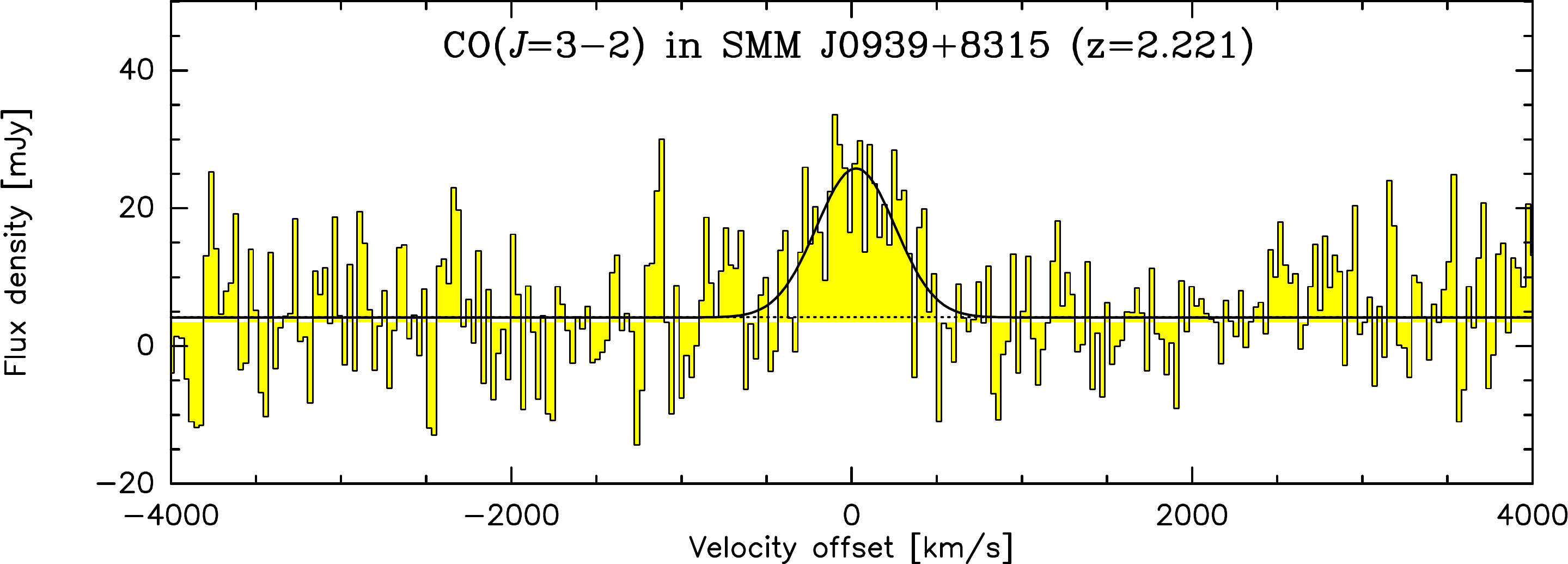}
\caption{Top Left: Continuum-subtracted moment-0 map of \CO line emission toward
the background SMG with $\sigma$\,=\,1.03\,Jy\,\,km\,\,s\pmOne\ beam\pmOne\ over a velocity range of $\Delta v$\,$\sim$\,514\,km\,\,s\pmOne. The beam size is 11$\farcs$5\,$\times$\,6\farcs2, at P.A.pdf.\,=\,$-$56\degr, as indicated in the bottom left corner.
Top Right: Velocity-integrated \CO line emission (red contours) overlaid on the SMA 1\,mm dust continuum (green contours and grayscale; H14), with an rms noise of $\sigma_\textrm{1\,mm}$\,=\,0.84 mJy beam\pmOne. The beam size of the SMA observations is 1\farcs4$ \times $1\farcs2, P.A. $-$34\degr, as shown
in the bottom left corner.
The central cross on each image corresponds to the same coordinates as in Figure~\ref{fig:cont}. The contour levels
in both images
start at $\pm$3$\sigma$, incrementing at
steps of $\pm$1$\sigma$.
Bottom:
Spectrum extracted at the peak position of CO line emission, with a spectral resolution of $\Delta v$ $\sim$\,29 km\,\,s\pmOne, and an rms of $\sigma_\textrm{ch}$\,=\,9.5 mJy beam\pmOne\ per channel. The
solid black line shows a Gaussian fit to the \CO line profile, where the velocity scale is relative to $z$\,=\,2.221.
\label{fig:mom0}}
\end{figure*}

We detect \CO line emission at $\sim$\,8$\sigma$ significance toward the background SMG SMM\,J0939 at $z$\,=\,2.221.
The lensing-magnified spatial extent of this SMG is $\sim$\,5$\arcsec$, as shown in the Submillimeter Array (SMA) 1\,mm dust continuum image in Figure~\ref{fig:mom0} (H14); as such,
the detected \CO line emission is spatially unresolved. We therefore extract the line profile (Figure~\ref{fig:mom0}) at the peak position of the unresolved
CO emission. Fitting a four-parameter single Gaussian to the spectrum yields a peak flux density of 21.61\,$\pm$\,2.66\,\,mJy, superimposed on a
continuum level of 4.15\,$\pm$\,0.48~mJy~beam\pmOne, and a line full width at half-maximum (FWHM) of 546\,$\pm$\,36\,\,km\,\,s\pmOne.  \par
We construct a velocity-integrated (0$^\textrm{th}$ moment) map of the \CO line
emission after subtracting continuum emission in the visibility plane. This results in a velocity-integrated \CO line flux of $I_\textrm{CO}$\,=\,12.6\,$\pm$\,2.0 Jy km\,\,s\pmOne\ over a velocity range of $\Delta v$\,$\sim$\,1420 km\,\,s\pmOne, the uncertainty does not include $\sim$\,15\% calibration
uncertainty. Our \CO line measurement confirms the redshift of SMM\,J0939, yielding $z$\,=\,2.2212\,$\pm$\,0.0010.

\section{Analysis}
\subsection{Lens Modelling} \label{sec:Lens}
To study the intrinsic properties of the background galaxy, we determine the magnification factor and the half-light radius of the dust region by performing
lens modeling on the SMA 1\,mm continuum data presented by H14 of this system.
Lens modeling is carried out in the visibility
({\it uv-}) plane using an updated version of the publicly available software {\sc uvmcmcfit} \citep{uvmcmcfit15a},
details of the parametric lens model can be found in \citet{Bussmann15a}.
The surface mass densities of the two lensing galaxies, 3C220.3 and its companion galaxy B, are described by singular isothermal ellipsoid profiles, and the source is assumed to have an
elliptical Gaussian profile. 
\begin{figure}[!tbpH]
\centering
\includegraphics[width=0.232\textwidth]{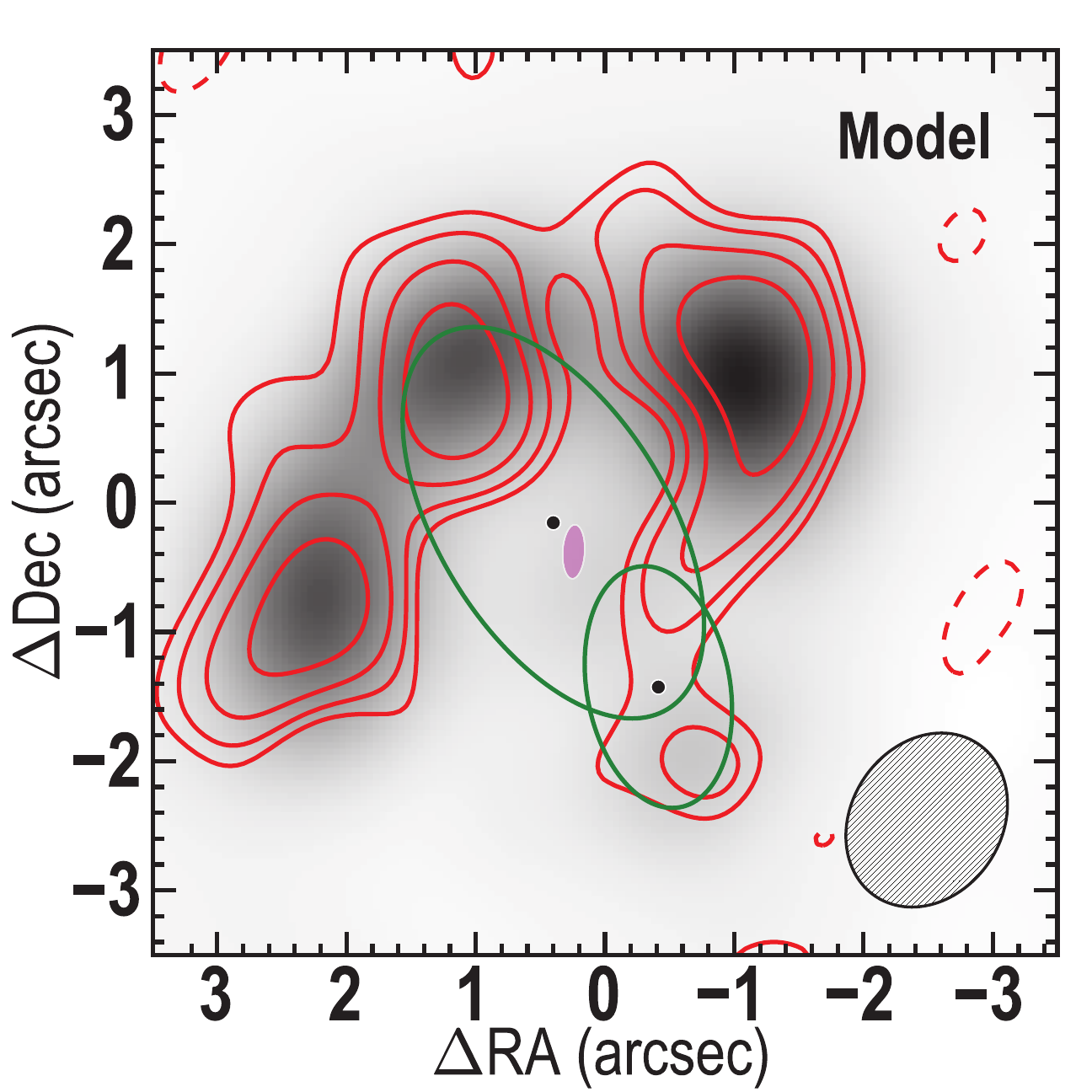}
\includegraphics[width=0.232\textwidth]{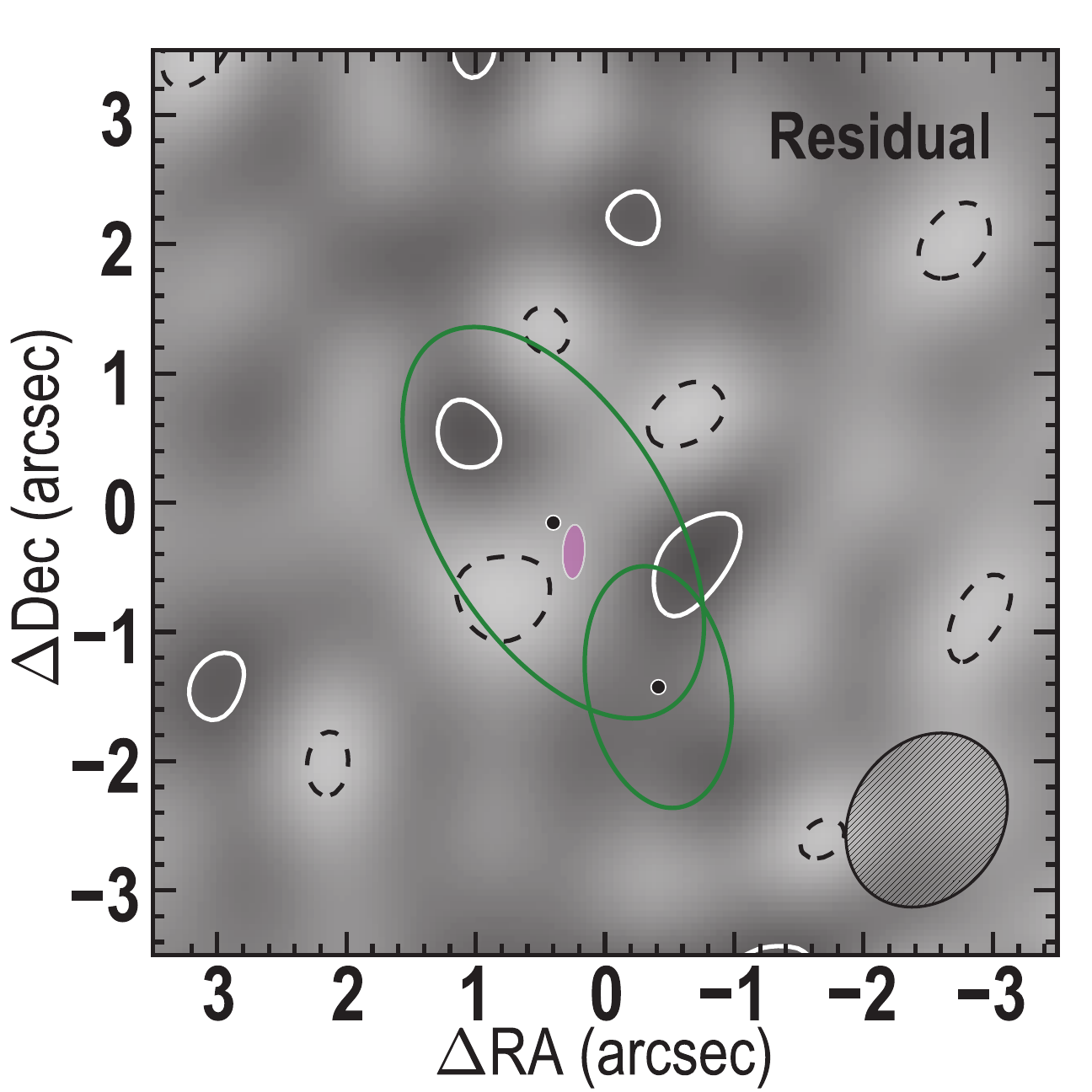}
\caption{Double-lens modeling of SMM\,J0939 using {\sc uvmcmcfit} on the SMA 1\,mm continuum data.
The contours start at $\pm$2$\sigma$, incrementing at
steps of $\pm$2$\sqrt{\rm 2}\sigma$ in both panels. Left: SMA 1\,mm continuum (red contours) overlaid on the best-fit model (grayscale image), assuming an elliptical Gaussian profile for the background SMG. The lenses are represented by the black dots, the half-light area of the background source is represented by the magenta ellipse, and the critical curves are represented by the green ellipses.
Right: Residual contours and image obtained by taking the Fourier transform of the difference between the SMA data and the best-fit model in the visibility plane. Solid (dashed) contours show the positive (negative) residuals.\label{fig:lens}}
\end{figure}

The resulting best-fit model as shown in Figure\,\,\ref{fig:lens} shows no significant bowls in the residual
image, and the knots (lensed emission) in the observed SMA data are reproduced well by the best-fit model.
 Our best-fit model yields a magnification
factor of $\mu_\textrm{L}$ = 10.13\,$\pm$\,1.38
and a half-light radius of $r_{s}$\,=\,0\farcs11 $\pm$ 0\farcs03, corresponding to $\sim$0.9 kpc at $z$\,=\,2.221.
All best-fit
parameters are listed in Table~\ref{tab:lensParam}.
\begin{deluxetable}{c l l r}[tbpH]
\tabletypesize{\scriptsize}
\tablecolumns{4}
\tablewidth{0pc}
\tablecaption{Lens modeling parameters and results}
\tablehead{
\multicolumn{3}{c}{Parameters} &
\colhead{Best-Fit Values} \vspace{0.05in}
\\ \cline{1-4} \vspace{-0.05in} \\
\multicolumn{4}{c}{Lens 0 (3C220.3)}
}
\startdata
Offset in RA        & $\Delta \alpha_{\rm lens0}$ & (\arcsec)   & 0.403 $\pm$ 0.026     \\
Offset in Dec        & $\Delta \delta_{\rm lens0}$ & (\arcsec)   & -0.181 $\pm$ 0.027    \\
Axial Ratio          & $q_{\rm lens0}$             &             & 0.446 $\pm$ 0.063     \\
Position angle       & $\phi_{\rm lens0}$          & (deg)       & 31.56 $\pm$ 4.15\phn  \\
Einstein radius      & $\theta_{\rm E0}$           & (\arcsec)   & 1.218 $\pm$ 0.010     \\
\cutinhead{Lens 1 (Companion galaxy B)}
Offset in RA         & $\Delta \alpha_{\rm lens1}$ & (\arcsec)   & -0.804 $\pm$ 0.034    \\
Offset in Dec        & $\Delta \delta_{\rm lens1}$ & (\arcsec)   & -1.243 $\pm$ 0.017    \\
Axial ratio          & $q_{\rm lens1}$             &             & 0.608 $\pm$ 0.138     \\
Position angle       & $\phi_{\rm lens1}$          & (deg)       & 14.2 $\pm$ 15.7\phn      \\
Einstein radius      & $\theta_{\rm E1}$           & (\arcsec)   & 0.745 $\pm$ 0.015     \\
\cutinhead{Source (SMM\,J0939)}
Offset in RA         & $\Delta \alpha_{\rm s}$     & (\arcsec)   & -0.163 $\pm$  0.035   \\
Offset in Dec        & $\Delta \delta_{\rm s}$     & (\arcsec)   & -0.193 $\pm$  0.048   \\
Axial ratio          & $q_{\rm s}$                 &             & 0.424 $\pm$ 0.237     \\
Position angle       & $\phi_{\rm s}$              & (deg)       & 174.34 $\pm$ 8.89\phn \\
Effective radius    & $r_{\rm s}$             & ($\arcsec$) & 0.106 $\pm$   0.033   \\
Magnification factor & $\mu_\textrm{L}$            &             & 10.13 $\pm$ 1.38\phn
\enddata
\label{tab:lensParam}
\tablecomments{All angular offsets are with respect to $\alpha$\,=\, 9$^{\rm h}$39$^{\rm m}$23\fs54, $\delta$\,=\,83\degr15\arcmin26\farcs10 (J2000). The corresponding masses within the Einstein radii of the galaxies 3C220.3 and its companion galaxy B are $M(\theta$\,\,$<$\,\,$\theta_\textrm{E})$\,=\,(4.86\,$\pm$\,0.08)\,$\times$\,10$^{11}$\,\,\Msun\ and $M(\theta$\,\,$<$\,\,$\theta_\textrm{E})$\,=\,(1.82\,$\pm$\,0.07)\,$\times$\,10$^{11}$\,\,\Msun, respectively.}
\end{deluxetable}

\subsection{SED Fitting} \label{sec:SED}
\subsubsection{3C220.3}\label{sec:SEDFg}
Synchrotron continuum emission from extended components of a radio galaxy decreases with increasing radio frequencies,
and the spectrum is commonly characterized by a power law distribution $S \propto \nu^{-\alpha}$, where the
spectral index $\alpha$ is $\gtrsim$ 0.5. While the contribution from extended components decreases, studies using
samples of radio galaxies have suggested that the flat/inverted-spectrum of the compact radio core component rises
and dominates the flux density at higher frequencies \citep{Kellermann81a,Begelman84a}. This has been observed in a FR-II galaxy at similar redshift --- 3C220.1 at $z$ = 0.610, where observations were carried out at the observed-frame frequency of $\sim$\,90 GHz \citep{Hardcastle08a}.

\begin{figure}[!tbph]
\centering
\includegraphics[width=0.5\textwidth]{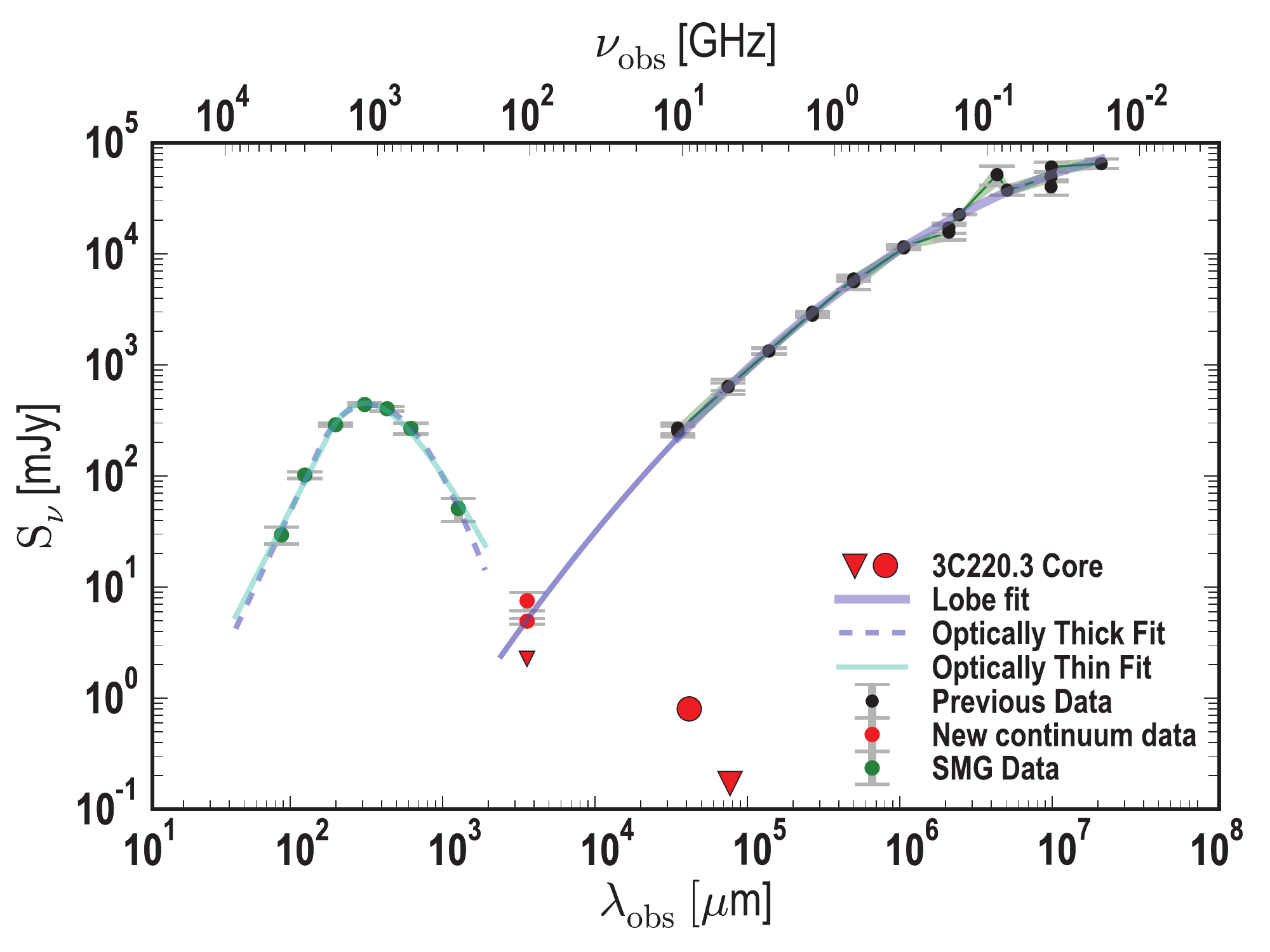}
\caption{SEDs of 3C220.3 (solid purple line) and SMM\,J0939 (dashed purple line and solid cyan line) including the new measurements presented in this paper.
The solid purple line corresponds to the parabolic function we
fit to the existing data associated with 3C220.3 (black dots; see Table \ref{tab:SEDdataRadio}).
The red dots at 104 GHz correspond to
our continuum measurements (integrated and peak, respectively), and the red triangles correspond to the upper limits on the radio core.
The dashed purple line and
the solid cyan line correspond to the best-fit optically thick and optically thin models of SMM\,J0939, respectively, using the photometric data from H14. \label{fig:SED}}
\end{figure}

Previously,  an upper
limit of $<$ 0.17 mJy at 4.6 GHz has been established by \citet{Mullin06a} on the core component of 3C220.3, and an unambiguous detection of 0.8 mJy at 9 GHz has been reported by H14, suggesting a substantially inverted spectrum of the core (Figure~\ref{fig:SED}).
Consequently, we may naively expect the integrated flux density in our continuum detection of $S_\textrm{104GHz}$\,=\,7.39\,$\pm$\,1.42\,\,mJy to be dominated by the unresolved core component of the foreground FR-II galaxy, which is at $z$\,=\,0.685.
However, the deconvolved spatial size of the source matching that in the resolved image (see Figure~\ref{fig:cont}) is
suggestive of a marginally resolved detection of the extended lobe components.
This is plausible given that the orientation of the synthesized beam in our observations is in alignment with the
axis along the
lobes of the radio galaxy, as shown in Figure~\ref{fig:cont}. We investigate this disparity by fitting models to
existing SED measurements as listed in Table \ref{tab:SEDdataRadio}, and extrapolating the fit to
estimate the flux density of the lobes at the frequency of our continuum measurement.
\begin{deluxetable}{rlrcc}[tbpH]
\tabletypesize{\scriptsize}
\tablecolumns{5}
\tablecaption{Continuum data of the lensing galaxy 3C220.3 and background SMG SMM\,J0939}
\tablehead{
\multicolumn{2}{c}{Wavelength} &
\multicolumn{2}{c}{Flux Density} &
\colhead{Instrument} \vspace{0.05in}
\\  \cline{1-5} \vspace{-0.05in} \\
\multicolumn{5}{c}{SMM\,J0939}
}
\startdata

70   & $\micron$ & 29.5 $\pm$ 5            & mJy & PACS      \\
100  & $\micron$ & 102 $\pm$ 7             & mJy & PACS     \\
160  & $\micron$ & 289 $\pm$ 9             & mJy & PACS   \\
250  & $\micron$ & \phn440 $\pm$ 15        & mJy & SPIRE    \\
350  & $\micron$ & \phn403 $\pm$ 20        & mJy & SPIRE   \\
500  & $\micron$ & \phn268 $\pm$ 30        & mJy & SPIRE   \\
1000 & $\micron$ & \phn\phn51 $\pm$ 14\tna & mJy & SMA     \vspace{0.05in} \\ \hline
 \cline{1-5} \vspace{-0.08in} \\
\multicolumn{2}{c}{Frequency} & \multicolumn{2}{c}{Flux Density} & \colhead{Reference} \vspace{-0.05in} \\
\cutinhead{3C220.3 Integrated (Core \& Lobes)}
    104.2 & GHz & 7.39 $\pm$ 1.42\tnb        & mJy & LR16 \\
    10.7  & GHz & 270 $\pm$ 30            & mJy & KP73       \\
    10.7  & GHz & 253 $\pm$ 28            & mJy & L80       \\
    5.0   & GHz & 640 $\pm$ 100           & mJy & K69       \\
    5.0   & GHz & 636 $\pm$ 50            & mJy & L80       \\
    2.7   & GHz & 1.33 $\pm$ 0.07         & Jy  & K69       \\
    2.7   & GHz & 1.34 $\pm$ 0.10         & Jy  & L80       \\
    1.4   & GHz & 2.95 $\pm$ 0.09         & Jy  & C98       \\
    1.4   & GHz & 2.99 $\pm$ 0.06         & Jy  & P66       \\
    1.4   & GHz & 2.80 $\pm$ 0.14         & Jy  & K69       \\
    1.4   & GHz & 2.89 $\pm$ 0.09         & Jy  & L80       \\
    0.75  & GHz & 5.94 $\pm$ 0.28         & Jy  & L80       \\
    0.75  & GHz & 5.94 $\pm$ 0.21         & Jy  & P66       \\
    0.75  & GHz & 5.60 $\pm$ 0.84         & Jy  & K69       \\
    352   & MHz & 11.3 $\pm$ 0.453        & Jy  & WENSS     \\
    352   & MHz & 11.6 $\pm$ 0.464        & Jy  & WENSS     \\
    178   & MHz & 15.7 $\pm$ 2.35         & Jy  & K69       \\
    178   & MHz & 17.1 $\pm$ 1.71         & Jy  & L80       \\
    152   & MHz & 22.6 $\pm$ 0.08         & Jy  & B85       \\
    152   & MHz & 22.5 $\pm$ 0.04         & Jy  & B85       \\
    86    & MHz & 51.6 $\pm$ 9.90         & Jy  & L80       \\
    73.8  & MHz & 37.5 $\pm$ 3.82         & Jy  & C07       \\
    38    & MHz & 49.6 $\pm$ 4.96         & Jy  & L80       \\
    38    & MHz & 40.2 $\pm$ 6.30         & Jy  & K69       \\
    37.8  & MHz & 60.7 $\pm$ 6.07         & Jy  & H95       \\
    17.8  & MHz & 64.9 $\pm$ 6.49         & Jy  & H95			\\
\cutinhead{3C220.3 (Core Only)}
    104.2 & GHz & $<$ 2.29\tnc 		      & mJy & LR16 \\
    9.0   & GHz & 0.80  $\pm$ 0.06    & mJy & H14       \\
    4.86  & GHz & $<$ 0.17            & mJy & M06       \\
\enddata
\label{tab:SEDdataRadio}
\tablecomments{Photometric data of SMM\,J0939 are from \citet{Haas14}.}
\tablenotetext{a}{Errors include calibration uncertainties}
\tablenotetext{b}{Integrated flux density. Peak flux density of the continuum emission is 4.93 $\pm$ 0.31 mJy beam\pmOne}
\tablenotetext{c}{Constraint from SED modeling}
\tablenotetext{$\dagger$}{www.astron.nl/wow/testcode.php?survey=1}
\tablerefs{
B85 = \citet{r2728};
C98 = \citet{r16};
C07 = \citet{r30};
H95 = \citet{r33-34};
H14 = \citet{Haas14};
K69 = \citet{r11-14-18-22-25-32};
KP73 = \citet{r9};
L80 = \citet{r10-13-15-19-20-26-29-31};
LR16 = this work;
M06 = \citet{Mullin06a};
P66 = \citet{r17-21};
WENSS = \citet{r23-24}$^\dagger$
}
\end{deluxetable}

Following Equation (1) in \citet{Cleary07a}, the fit to the lobe emission can be expressed as a parabolic function:
\begin{equation}
\log F_{\nu}^\mathrm{lobe} (\nu) \propto - \beta\ (\log\ \nu - \log \nu_\mathrm{t})^2  + \log (\exp({\frac{\nu}{\nu_c^\mathrm{lobe}}}))
\end{equation}
where $F_{\nu}^\mathrm{lobe}$ is the flux density of the lobes, $\beta$ is a parameter representing the bending
of the parabola, $\nu_\mathrm{t}$ is the frequency at which the optical depth of the synchrotron emitting plasma reaches
unity, and $\nu_c^{\rm lobe}$ is the frequency corresponding to the cutoff energy of the lobe plasma energy
distribution.
The extrapolated flux density at 104\,\,GHz is consistent with the peak flux density of our continuum
measurement (Figure~\ref{fig:SED}). The 9$\sigma$ detection of the continuum thereby suggests a
dominant contribution from the lobes, and that the peak flux density is not dominated by emission toward
the core. Moreover, the peak position of the 104\,\,GHz continuum is
centered toward the brighter northern lobe (Figure~\ref{fig:cont}), which further supports our argument.
Consequently, a conservative upper limit of $S_\nu$\,$<$\,4.93 mJy on the core emission can be established using the measured peak flux density. Yet, by considering the
difference between the integrated flux density from our measurement and the flux density from an extrapolation of the model (see Figure~\ref{fig:SED}; $S_\textrm{104GHz, fit}$~=~5.10 mJy), we establish a more stringent constrain on this upper limit of $S_\nu$\,$<$\,2.29 mJy. We did not
 extrapolate the core measurements to the frequency of our continuum, as previous measurements of the core are
 taken
 across different epochs, and the core may be time-variable.
\par
Studies by \citet{Meisenheimer89a} and \citet{Hardcastle08a} have suggested that spectra of hotspots are flat up to optical frequencies, where some exhibit spectral steepening in cm and mm wavelengths (\eg 3C123). At the resolution of our observations, it remains unclear whether the measured flux density is dominated by emission from the compact hotspots or that from the surrounding diffuse lobe components.

\subsubsection{SMM\,J0939+8315} \label{sec:SEDBg}
To constrain the dust and gas properties in the ISM of SMM\,J0939, we perform SED fitting to the
photometric data obtained with {\it Herschel}/PACS and SPIRE, at wavelengths
between observed-frame 70\,\micron\,$-$\,500\,\micron, and the interferometric data obtained with the SMA at 1\,mm (H14). We use the publicly
available software {\sc mbb\_emcee}\footnote{\url{https://github.com/aconley/mbb\_emcee}} to perform the SED fitting; the code uses an affine-invariant Markov chain Monte
Carlo (MCMC) approach, and further details of the code are given by \citet{Riechers13a} and \citet{Dowell14a}. \par
The
functional form of the fit comprises a single-temperature, modified blackbody function joined to a $B_{\lambda} \propto \lambda^\alpha
$ power law on the blue
side of the SED.
We fit both optically thick and optically thin models. In the optically thick case, the wavelength $
\lambda_0$\,=\,${c}/{\nu_0}$ is an additional parameter representing the rest-frame wavelength at which the optical
depth $\tau_{\nu} =$ ($\nu$/$\nu_0$)$^\beta$ reaches unity. Thus, the functional form of the modified blackbody
in the optically thick regime is as follows:
\begin{equation}
\rm B_{\lambda} \propto \frac{(1-exp^{-(\frac{\lambda_0 (1+z)}{\lambda})^{\beta}})(\frac{c}{\lambda})^3}
{exp^{\frac{hc}{\lambda\rm{kT/(1+z)} } }-1}
\end{equation}
and in the optically thin regime, the functional form reduces to:
\begin{equation}
\rm B_{\lambda} \propto \frac{(\frac{c}{\lambda})^{\beta+3}}{exp^{\frac{hc}{\lambda\rm{kT/(1+z)}}}-1}
\end{equation}
where $T$ is the rest-frame cold dust temperature, $\beta$ is the dust emissivity index
, and $\alpha$ is the mid-infrared power law spectral index. The overall fit is normalized using the observed-frame 500
$\micron$ flux density, hence this becomes an additional parameter ($f_{\rm norm,\ 500\mu m}$) in the fit. For both models, we impose an upper limit of 60 K on the observed-frame dust temperature ($T/(1+z)$), and an upper limit of 2.2 on
$\beta$. For the optically thick model, we impose an additional upper limit of 3000\,\micron\ on $\lambda_0 (1+z)$.

\begin{deluxetable}{lccc}[tbpH]
\tabletypesize{\scriptsize}
\tablecolumns{4}
\tablecaption{SED fitting results}
\tablehead{
\multicolumn{2}{c}{Parameters}      &
\colhead{Optically Thick} &
\colhead{Optically Thin}
}
\startdata
$\chi^2$           &           & 2.25                     & 5.31                    \\
D.O.F              &           & 2                        & 3                       \\
$T$                & (K)           & 63.1$^{+1.1}_{-1.3}$     & 52.0$^{+1.3}_{-1.2}$    \\
$\beta$            &           & 1.9$^{+0.6}_{-0.5}$      & 0.7$^{+0.2}_{-0.3}$     \\
$\alpha$           &           & 2.9$^{+0.3}_{-0.4}$      & 2.8$^{+0.2}_{-0.2}$     \\
$\lambda_0$\tna    & ($\micron$)          & 248.7$^{+86.0}_{-123.8}$ & ---             \\
$\lambda_{\rm peak}$ \tnb  & ($\micron$)   & 254.7$^{+6.2}_{-6.1}$    & 301.4$^{+29.0}_{-30.1}$ \\
$f_{\rm norm,\ 500\mu m}$\tnc & (mJy) & 267.4$^{+16.7}_{-16.3}$  & 244.3$^{+15.3}_{-15.3}$ \\
$L_{\rm IR}$\tnd        & (10$^{12}$ \Lsun)      & 88.5$^{+2.6}_{-2.6}$     & 89.2$^{+2.5}_{2.5}$     \\
$M_{\rm dust}$\tne         & (10$^8$ M$_\odot$)   & 50.5$^{+20.4}_{-20.2}$   & 25.7$^{+3.9}_{-5.5}$
\enddata
\label{tab:mbb}
\tablenotetext{a}{Rest-frame wavelength where $\tau_\nu$\,=\,1}
\tablenotetext{b}{Observed-frame wavelength of the SED peak}
\tablenotetext{c}{Observed-frame flux density at 500 $\micron$}
\tablenotetext{d}{Rest-frame 8-1000 $\micron$ luminosity}
\tablenotetext{e}{Derived assuming a standard absorption mass coefficient $\kappa$=2.64 m$^2$ kg$^{-1}$ at $\lambda$=125.0 $\micron$ \citep{Dunne03a}}
\tablecomments{Errors reported here are $\pm$1$\sigma$. $L_{\rm IR}$ and $M_{\rm d}$ are reported prior to lensing correction.}
\end{deluxetable}

The best-fit values in both regimes are listed in Table \ref{tab:mbb}, and the correlation plots are available in the Appendix. The best-fit solution of optically thin
models corresponds to $\chi^2$ = 5.31 with 3 degrees of freedom, whereas that of optically thick models
corresponds to $\chi^2$ = 2.25 with 2 degrees of freedom, suggesting a better fit than in the optically thin
case. In the subsequent analysis, we employ the inferred values from the optically thick model.
The best-fit solution yields a far-infrared luminosity (rest-frame 42.5$-$122.5\micron) of $L_\textrm{FIR}$ = 53.3$^{+1.1}_{-1.1}$\,$\times$\,10$^{12}$\,\Lsun, and a total infrared (IR; rest-frame 8$-$1000 \micron) luminosity of $L_\textrm{IR}$ = 88.5$^{+2.6}_{-2.6}$\,$\times$\,10$
^{12}$\,\Lsun
\footnote{
Owning to the positive K-correction blue-ward of the dust peak, in which
the foreground radio galaxy contributes a non-negligible amount to the MIR luminosity, we do not fit for a separate AGN component.
Instead, we adopt a power-law to account for the MIR excess, which allows us to estimate the IR
luminosity \citep[\eg][]{Casey12a, Riechers13a, Kirkpatrick15a}. }.
Assuming a dust absorption coefficient of $\kappa_{\nu}$ = 2.64\,\,m$^2$\,\,kg\pmOne\ at 125.0\,\,$
\micron$ \citep{Dunne03a}, we find a dust mass of $M_\textrm{dust}$ = 50.5$^{20.4}_{-20.2}\times$10$^8$\,\,\Msun{\bf ;} the uncertainties do not include those in the dust absorption coefficient ($\kappa_{\nu}$). These properties are derived based on the SED fitting to the photometric data, \ie prior
to lensing correction. We note that the dust mass is weakly constrained owing to the dearth of data in the rest-frame FIR waveband. As such, we investigate how the dust mass would be affected by fitting additional optically thick models with an upper limit of
 $\beta$ adjusted from 2.2 to 3.0. While the difference in each best-fit parameter between this scenario and the previous models (with an upper limit of $\beta$ = 2.2) is within 3\%, we find that the dust mass inferred from this best-fit SED model is boosted by a factor of $\sim$\,2.
\subsection{Physical Properties of the ISM in SMM\,J0939}

\subsubsection{Molecular Gas Mass} \label{sec:gas}
While the ground state CO transition line traces the cold molecular gas in the ISM
\citep*[\eg][]{Wilson70a,Downes98a}, transition lines of higher rotational states ($J$ $>$ 1) are frequently observed in high-redshift sources as the
 ground state transition line is redshifted to lower frequencies that can only be observed with traditional radio telescopes
 \citep{Carilli13a}. Consequently, assumptions on the CO excitation conditions are required to derive the molecular gas mass using the $M$(H$_\textrm{2}$)-to-$L^{\prime}_\textrm{CO}$
 conversion factor ($\alpha_\mathrm{CO}$) when extrapolating from higher-$J$ CO lines. \par
Recent observations in high-redshift quasar hosts suggest that the ratio
   is $R_\textrm{31}$\,$\sim$\,1 \citep{Riechers06a, Riechers11a}. In the case of high-redshift type-2 quasars, \citet{Riechers11a} report a brightness temperature ratio of $R_\textrm{31}$ = 1.00\,$\pm$\,0.10 for IRAS F10214+4724 (hereafter F10214), which is currently the only known type-2 quasar with both \CO and CO($J$\,=\,1\,\rarr\,0) line measurements.
Here, we derive the molecular gas mass
assuming thermalized excitation of CO, as SMM\,J0939 is
postulated to be hosting a type-2 quasar (H14). \par
We calculate the CO($J$\,=\,1\rarr\,0) line luminosity using a standard relation
\citep[\eg][]{Solomon05a}
and assuming a conversion factor of $\alpha_\textrm{CO}$\,= 0.8\,\Msun\,(\LpU)\pmOne\ based on empirical relations from local ULIRGs, which is typically
adopted for SMGs \citep[\eg][]{Tacconi06a,Tacconi08a,Bothwell13a}.
This corresponds to \Lp = (3.42\,$\pm$\,0.71)\,$\times$\,10$^{10}$\,(10.1/$\mu_\textrm{L}$) \LpU; hence the inferred total molecular gas mass is $M_\textrm{gas}$ = (2.74\,$\pm$\,0.57)\,$\times$\,\,10$^{10}$\,\Msun\, after correcting for lensing magnification. This results in a gas-to-dust
ratio of $f_\textrm{gas-dust}$\,=\,$M_\textrm{gas}/M_\textrm{dust}$\,=\,55\,$\pm$\,24.
This is in good agreement  with the
values found for other SMGs \citep{Coppin08a,Micha10a,Riechers11c}.

\subsubsection{Star Formation Rate \& Star Formation Efficiency}
We derive the SFR using the lensing-corrected far-infrared luminosity
assuming
that the dominant heating source of cold-dust is young and massive stars, and
that a contribution from the
dust-enshrouded AGN is negligible.
This assumption stems from the results of recent studies using various approaches, such as spectral decomposition techniques and correlation between far-infrared luminosity and other tracers of star-formation, suggesting that far-infrared emission dominantly originates from star-formation in host galaxies, even in the most energetic QSOs \citep[\eg][]{Netzer07a, Mullaney11a, Harrison15a}.

Using the \citet{Kennicutt98a} relation and adopting a \citet{Chabrier03a}
stellar initial mass (IMF) function, we find a
SFR$_\textrm{FIR}$\,=\,526\,$\pm$\,73 $M_
\odot$ yr\pmOne.
The starburst in SMM\,J0939 can be maintained at its
current rate for a time that can be approximated by the gas depletion timescale, $\tau_\textrm{depl}$\,=\,$M_\textrm{gas}$/SFR, which assumes no replenishment of gas and feedbacks.
This corresponds to $\tau_\textrm{depl}$\,=\,52\,$\pm$\,8 Myr, which is in good agreement with those found in other SMGs \citep[\eg][]{Greve05a}.

The SFR per unit mass of molecular gas is commonly taken as a
measure of the star formation efficiency.
We compute this ratio using the far-infrared
and CO luminosities. The derived SFE is therefore independent of the magnification factor, the CO luminosity to gas mass conversion factor ($\alpha_\textrm{CO}$), and the
IMF. This, however, assumes that differential lensing between the CO and far-infrared emission is negligible.
The resulting ratio is SFE$_\textrm{FIR}$\,=\,154\,$\pm$\,25\,\,\Lsun\,(\LpU)$^{-1}$, this is comparable
to those found in ``typical" SMGs \citep{Greve05a,Tacconi06a,Riechers11c}.

We compute the surface densities by dividing half the SFR and gas mass by the area subtended by the half-light
radius \citep[\eg][]{Genzel10a, Harrison15a}, yielding
$\Sigma_\textrm{SF}$ = 106  \Msun~yr\pmOne~kpc$^{-2}$ and $\Sigma_\textrm{gas}$ = 5.48\,$\times$\,10$^9$ \Msun~kpc$^{-2}$, respectively.
These results are in good agreement with values typical for SMGs \citep{Tacconi06a, Hodge15a}. 
The inferred surface densities of SMM\,J0939 follow a universal Schmidt-Kennicutt relation between the star formation rate
surface density and the molecular gas surface density: $\Sigma_\textrm{SF}$ = 9.3\,($\pm$\,2)\,$\times$\,10$^{-5}$ ($M_\textrm{gas}$/2$\pi R_\textrm{1/2}^2)^{1.71(\pm\,0.05)}$, which was derived using a sample consisting of local star-forming galaxies and high-redshift
galaxies
out to $z$\,$\sim$\,2.5, and assuming a Chabrier IMF \citep{B07a}.

\subsubsection{Physical Size and Dynamical Mass}
Our lens model suggests a half-light radius of $r_s$\,$\sim$\,1\,kpc for the dust-emitting region in SMM\,J0939. This is comparable to the half-light radii found in other SMGs with high resolution imaging. Similar sizes have been reported by \citet{Bussmann13a}, who find typical radii of 1.5 kpc for a sample of {\it Herschel}$-$selected lensed SMGs with $S_{500\micron}\,>$\,100 mJy.
Also, \citet{Simpson15a} report a radial extent of 1.2 kpc for a sample of un-lensed SMGs with $S_{850\micron}$ = 8$-$16 mJy.

We estimate the dynamical mass of SMM\,J0939 using our \CO line measurement and assuming that the molecular gas is virialized. With this assumption, we use an isotropic virial estimator \citep[\eg][]{Engel10a}, with the FWHM of the \CO line profile and the half-light radius from our lens model for $R_\textrm{eff}$, assuming that the dust emission traces the same emitting region as the CO.
We find a dynamical mass of $M_\textrm{dyn}$ = (7.84\,$\pm$\,2.84)\,$\times$\,10$^{10}$\,\Msun, and a gas-to-dynamical mass fraction of $f_\textrm{gas-to-dyn}$ = 0.35\,$\pm$\,0.14, consistent with those of other SMGs \citep{Tacconi06a}.  We note that a derived dynamical mass based on this assumption is likely to be biased towards low values, as the CO-emitting region can be apparently more extended than the dust emitting region
 due to the low dust optical depth at larger radii.
This is supported by recent studies, in which CO source sizes ranging from $\sim$4$-$20 kpc have been found, which are larger than typical dust continuum sizes \citep{Tacconi06a, Riechers11c, Ivison11a, Hodge13a, Hodge15a}.

\section{Discussion And Conclusions}
We present the detection of \CO line emission toward SMM\,J0939+8315, a strongly-lensed SMG that is hosting a type-2 quasar, refining the redshift
to $z$\,=\,2.2212\,$\pm$\,0.0010. The underlying continuum is detected at $\sim$\,9$\sigma$ significance, where the flux density is likely dominated by emission from the lobes and hotspots of the foreground radio galaxy 3C220.3.

The detection of CO in SMM\,J0939
implies a CO luminosity of \Lp\ = (3.4\,$\pm$\,0.7)\,$\times$\,10$^{10}$\,(10.1/$\mu_\textrm{L}$) \LpU, corresponding to a gas mass of
$M_{\rm gas}$\,=\,(2.7$\pm$0.6\,$\times$\,10$^{10}$(10.1/$\mu_\textrm{L}$ $\Msun$).
This 
suggests the presence of a massive
molecular gas reservoir that fuels the star formation activity taking place at a rate of $\sim$\,526 \Msun~yr\pmOne. If the star forming activity continues at the current rate, the gas reservoir will be depleted within $\lesssim$~52 Myr,
which is consistent with the short timescales found in other SMGs \citep{Greve05a}. The derived intrinsic properties of SMM\,J0939 are evident of ongoing rapid star formation; this is in good agreement with the current conjecture that SMGs are a
population of high-redshift galaxies that build up the bulk of stellar mass in present-day galaxies, thus play an important role
in galaxy formation and evolution \citep[\eg ][]{Dickinson03a}.

We
compare our findings for SMM\,J0939
with a sample of typically unlensed or only weakly magnified, 850~$\micron$$-$selected SMGs \citep[][hereafter B13]{Bothwell13a}.
Their properties are listed in Table~\ref{tab:compareSMG}, showing that SMM\,J0939 has properties similar
to other SMGs studied to date.
The gas mass in SMM\,J0939 is slightly lower than the median in the B13 sample, but we cannot rule out the possibility that this difference is due to the different assumptions made for the gas excitation conditions.
The gas properties (CO luminosity and gas mass) of the SMGs in the B13 sample are derived based on the assumption of typical excitation conditions found from CO spectral line energy distribution (SLED) modelling of the sample average, which the authors find to be very similar to those of the cosmic Eyelash. Thus, we additionally compare SMM\,J0939 in more detail to two other well-studied, strongly-lensed SMGs
with comparably high apparent submillimeter fluxes
found at similar redshifts --- HLSW-01 and the cosmic Eyelash.
The properties of these sources are derived using similar approaches to those employed in this paper.

As shown in Table~\ref{tab:compareSMG}, while the cosmic Eyelash has the least amount of molecular gas, as well as
the
longest gas depletion timescale, the overall gas and dust properties of SMM\,J0939 fall between those of HLSW-01 and the
cosmic
Eyelash. Such distinction is likely a result of our selection bias: while these sources appear similarly bright at 250~\micron, the lensing magnification
varies by a factor of $\sim$\,3. In particular, with the cosmic Eyelash having the highest lensing magnification among the
three, this
intrinsically fainter, and less gas-rich SMG appears notably bright
at 250\,\micron, while its CO line and IR luminosities are lower than those of most SMGs studied to date.
While lensing can probe sources
of various intrinsic properties, we find that
SMM\,J0939 is consistent with the ``typical" SMG population, with its intrinsic CO line luminosity, IR luminosity,
dust mass, SFR, SFE, depletion timescale, and gas mass fraction comparable to those found in ``typical" SMGs studied to date.

Since SMM\,J0939 also hosts a type-2 quasar, we compare its properties
against those of eight CO-detected obscured AGNs at $z$ = 1.6\,$-$\,2.8 \citep[][and references
therein]{Polletta11a}.
Among these obscured quasars, F10214 has the lowest molecular gas mass as well as SFR. The fact that the gas mass of F10214 in their compilation was derived using \CO line emission \citep{Solomon05a} has a minor effect on the resulting low gas mass; \citet{Riechers11a} report
a similarly low gas mass derived using their CO($J$\,=\,1\,\rarr\,0) line emission.
With F10214 being the most strongly-lensed high-redshift type-2 quasar \citep[$\mu_{\textrm L}$ = 17; ][]{Solomon05a}\footnote{Note that \citet{Deane13a} suggest a magnification factor of $\mu_{\textrm L}$ = 6\,$\pm$\,1.5 for the CO emission in F10214.}, its exceptionally low molecular gas mass and SFR is evident that this source lies on the low end of the CO and IR luminosity distributions of the population.
In contrast to what was found for F10214, we find that the properties
(\eg FWHM of the CO line profile, M$_\textrm{gas}$, and SFE) of SMM\,J0939 are similar to the statistical means, except for the SFR, which is lower by a
factor of $\sim$\,1.5, but is nevertheless consistent within the measurement uncertainties. We thus find that the gas mass in SMM\,J0939 is consistent with other type-2 quasars. 

While the gas mass in SMM\,J0939 is slightly lower than the median of SMGs in the B13 sample, we find it to be consistent with other type-2 quasars, suggesting a possible scenario in which a significant fraction of the gas has already been converted into stars and used for fueling the quasar.
As such, given the presence of an obscured quasar and the overlap of properties
with both type-2 quasar and SMG populations,
a physical interpretation might be that SMM\,J0939 is transitioning from a short, star-bursting phase to an unobscured quasar phase, consistent with the proposed evolutionary link model between dusty starbursts and quasars \citep[\eg][]{Sanders88,Coppin08a,Simpson12a}.

\begin{acknowledgments}
We thank the referee for providing constructive comments to improve this manuscript.
We thank Shane Bussmann for providing the code {\sc uvmcmcfit} for lens modeling, sharing the SMA and VLA data, and for helpful discussions; Alex Conley for providing the code {\sc mbb\_emcee} for SED fitting.
Support for CARMA construction was derived from the
Gordon and Betty Moore Foundation, the Kenneth T. and Eileen
L. Norris Foundation, the James S. McDonnell Foundation, the
Associates of the California Institute of Technology, the University
of Chicago, the states of Illinois, California, and Maryland,
and the National Science Foundation.
Ongoing CARMA development
and operations are supported by the National Science
Foundation under a cooperative agreement, and by the CARMA
consortium universities.

Facilities: CARMA
\end{acknowledgments}


\clearpage
\begin{turnpage}
\begin{deluxetable}{l l c c c cc c cc c c c c}[tbpH]
\tablecaption{Comparison of SMM J0939 with SMGs and type-2 QSOs at $z\sim$\,2. \label{tab:compareSMG}}
\tabletypesize{\scriptsize}
\tablecolumns{14}
\tablehead{
\multicolumn{2}{c}{}       &
\colhead{}                     &
\colhead{SMM J0939}  &
\colhead{}                     &
\multicolumn{2}{c}{HLSW-01}    &
\colhead{}                     &
\multicolumn{2}{c}{Cosmic Eyelash} &
\colhead{}                     &
\colhead{SMGs}            &
\colhead{}                     &
\colhead{Type-2 QSOs}    \\
\cline{4-4} \cline{6-7} \cline{9-10} \cline{12-12} \cline{14-14} \\
\vspace{-1.4em}\\
\colhead{Quantity}        &
\colhead{Unit}               &
\colhead{}                     &
\colhead{}                     &
\colhead{}                     &
\colhead{}                     &
\colhead{Reference}                     &
\colhead{}                     &
\colhead{}                     &
\colhead{Reference} &
\colhead{}                     &
\colhead{}                     &
\colhead{}                     &
\colhead{}
}
\startdata
$z$                       &                     && 2.221             && 2.957                 &  R11 && 2.326              & S10  && 2.2\tna && 2.27\,$\pm$\,0.32\tna \\
$\mu_{\rm L}$             &                     && 10.1 $\pm$ 1.4    && 10.9 $\pm$ 0.7       &  G11 && 37.5 $\pm$ 4.5     & S11  && -- && -- \\
$S_{\rm 250}$             & mJy                 && 440 $\pm$15\tnb   && 425 $\pm$ 10          &  C11 && 366 $\pm$ 55       & I10  && -- && -- \\
$I_\textrm{CO(3-2)}$ 
      & Jy km s$^{-1}$      && 12.6 $\pm$ 2.0    && 9.7 $\pm$ 0.5         &  R11 && 13.2 $\pm$ 0.1     & D11 && -- && --  \\
$\Delta v_{\rm FWHM}$ & km s$^{-1}$         && 546 $\pm$ 36\tnc  && 350 $\pm$ 25\tnc          &  R11 && $\lesssim$ 800\tnc\tnd & D11 && 550$\pm$90\tne && 450\,$\pm$\,180\tnf \\
\Lp                       & 10$^{10}$ \LpU      && 3.4 $\pm$ 0.7 && 4.2 $\pm$ 0.4         &  R11 && 1.7 $\pm$ 0.2      & D11 && 5.2 $\pm$1.0 && 3.0\tne  \\
$M_{\rm gas}$\tng             & 10$^{10}$ \Msun     && 2.7 $\pm$ 0.6 && 3.3 $\pm$ 0.3         &  R11 && 1.6 $\pm$ 0.1      & S10 && 4.2$\pm$0.8 && 2.4\,$\pm$\,1.4 \\
$L_{\rm FIR}$              & 10$^{12}$ \Lsun     && 5.3 $\pm$ 0.7  && 11.0 $\pm$ 0.9         &  C11 && 1.8 $\pm$ 0.2      & I10 && 6.0$\pm$0.6\tnh && 2.8\tni \\
$M_{\rm dust}$            & 10$^8$ \Msun        && 5.2 $\pm$ 2.1 && 1 $-$ 5.2                   & R11                 && $\sim$ 4.0        & I10 && 5.4$\pm$1.5\tni\tnj  && \nodata \\
SFR$_{\rm FIR}$\tnk
 & \Msun~yr$^{-1}$     && 526 $\pm$ 73  && 1430 $\pm$ 160    &  C11 && $\sim$ 235     & I10 &&  600$\pm$60\tni && 855 \,$\pm$\,480\\
$\tau_{\rm depl}$
& Myr                 && 52 $\pm$ 8    && 23 $\pm$ 3        &  R11 && 68\tni             & LR16 && 70$\pm$15\tni && 35\tni \\
$f_{\rm gas-dust}$    &                     && 55 $\pm$ 24       && 60 $-$ 330            &  R11 && $\sim$ 40          & I10 && 78$\pm$26\tni && \nodata \\
SFE                   & \Lsun\ (\LpU)\pmOne && 256 $\pm$ 41     && 340 $\pm$ 40          &  R11 && 135 $\pm$ 20\tni   & LR16 && 182 $\pm$ 38\tni && 347\,$\pm$\,268 \\
$M_{\rm dyn}$             & 10$^{10}$ \Msun     && 7.8 $\pm$ 2.8     && 3.7 $\pm$ 1.8\tni\tnl &  LR16 && 6.0$\pm$0.5      & S11 && 7.2$\pm$1.3 && \nodata \\
$f_{\rm gas-dyn}$         &                     && 0.4 $\pm$ 0.1     && 0.9\tni               &  LR16 && 0.6 $\pm$ 0.1      & S11 &&  0.6$\pm$0.2 && \nodata\\
\enddata
\tablecomments{Properties of SMGs and type-2 QSOs are based on the results from B13 and \citet{Polletta11a}, respectively.
Gas mass
is estimated based on the lowest-$J$ CO line measurements available. Thermalized excitation (\ie $R_{\rm 31}$\,=\,1) has been assumed for type-2 QSOs and
the excitation conditions for SMGs are based on CO SLED modeling \citep[see][]{Bothwell13a}.
 Values listed from row 6 onwards are lensing-corrected, and the errors quoted for SMM\,J0939 includes uncertainties in $\mu_{\rm L}$.}
\tablenotetext{a}{Statistical average in the sample}
\tablenotetext{b}{H14}
\tablenotetext{c}{\CO}
\tablenotetext{d}{Estimated from Figure 1 in D11}
\tablenotetext{e}{Based on \CO and \rot{4}{3} line observations}
\tablenotetext{f}{Based on \rot{2}{1}, \rot{3}{2}, and \rot{4}{3} line observations}
\tablenotetext{g}{$\alpha_{\rm CO}$\,=\,0.8\,\,\Msun\,(\LpU)\pmOne}
\tablenotetext{h}{Inferred from radio continuum measurements via the FIR-radio correlation (B13)}
\tablenotetext{i}{Derived from the reported values}
\tablenotetext{j}{Using $S_{\rm 850\micron}$ and optically thin, Rayleigh-Jeans approximation \citep{Scoville13a}}
\tablenotetext{k}{Chabrier IMF}
\tablenotetext{l}{Using the physical size of CO($J$ = 5 \rarr\ 4) emission (G11)}
\tablerefs{
C11 = \citet{Conley11a};
D11 = \citet{Danielson11a};
G11 = \citet{Gavazzi11a};
I10 = \citet{Ivison10c};
LR16 = this work;
R11 = \citet{Riechers11b};
S11 = \citet{Swinbank11a};
S10 = \citet{Swinbank2010a}}
\end{deluxetable}

\clearpage
\end{turnpage}

\begin{figure}[!tbph]
\centering
\hspace{-1.1cm}
\includegraphics[width=0.55\textwidth]{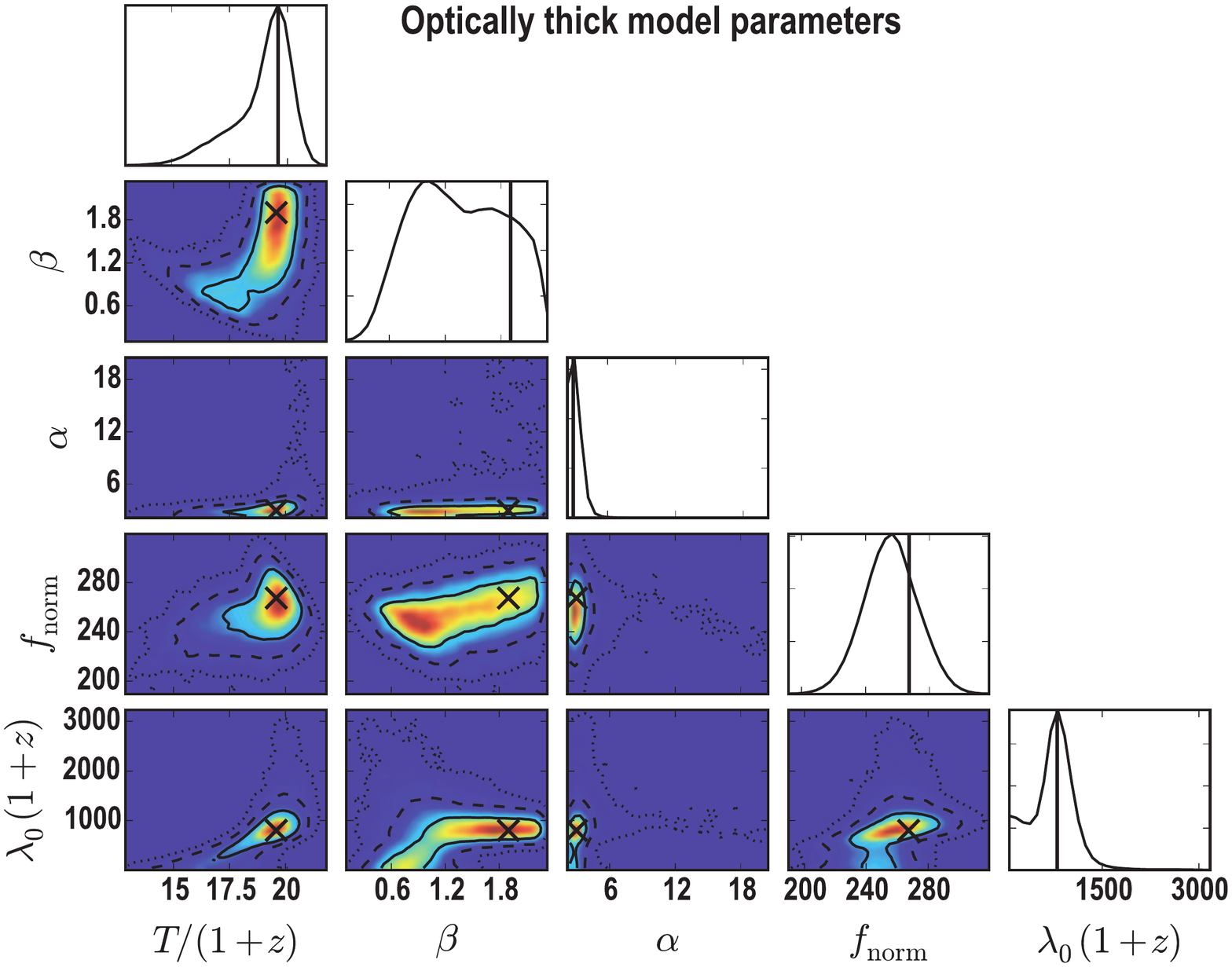}
\hspace{-1cm}
\includegraphics[width=0.55\textwidth]{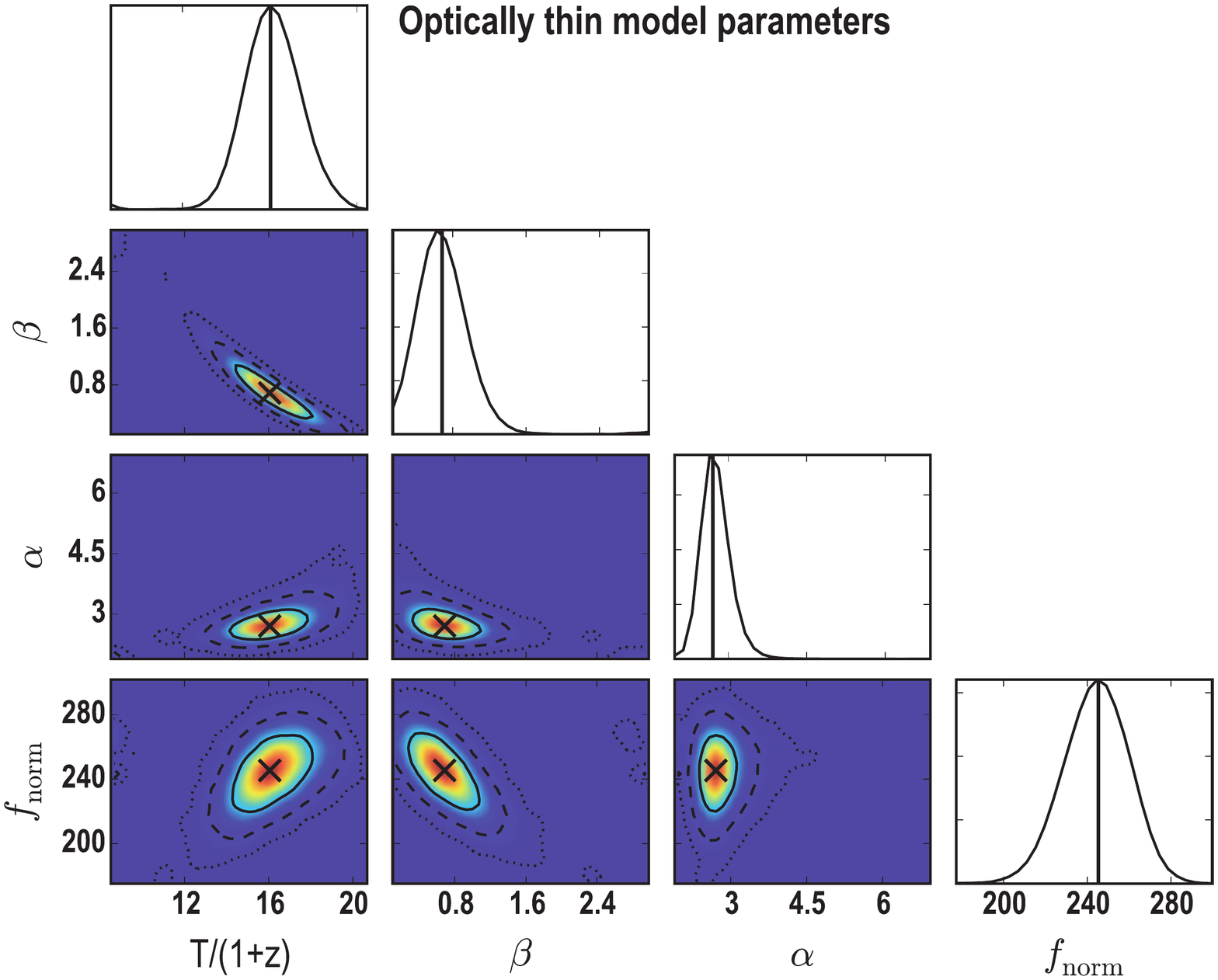}
\hspace{-1.1cm}
\caption{Correlation plots from the fitted SEDs, using optically thick
models (left), and optically thin models (right). Marginalized posterior probability
distributions of each
parameter are plotted along the diagonals, where the solid black vertical lines indicate the set of parameter values that maximizes the likelihood. The joint PDFs between parameters are plotted as 2D histograms on the off-diagonals, where the black crosses denote the
locations of the most likely solution in the parameter space. The solid, dashed, and dotted lines correspond to 1$\sigma$, 2$\sigma$, and $3\sigma$, respectively.
\label{fig:sedlikelihood}}
\end{figure}

\appendix
We perform SED fitting to the (sub)-mm data of SMM\,J0939 using {\sc mbb\_emcee} (see \S \ref{sec:SEDBg}). This code uses MCMC to sample the parameter spaces, the total IR luminosity and dust mass are then derived from the set of parameters that maximizes the likelihood.
We show in Figure~\ref{fig:sedlikelihood} the resulting correlation plots of each pair of parameters (off-diagonals) as well as the marginalized PDFs of each parameter (diagonals) for the fitted optically thick and optically thin models.
The parameter values corresponding to the most likely solution are denoted as black crosses in the joint probability distribution plots, and vertical lines in the marginalized PDF plots. \par

The PDFs of the parameters in the optically thin models are shown in the right panel of Figure~\ref{fig:sedlikelihood}, the dust emissivity index ($\beta$) is highly anti-correlated with the observed-frame dust temperature ($T/(1+z)$), and the marginalized PDFs of all parameters are well-approximated by Gaussians --- the best-fit solution is prominent.
We note that the best-fit emissivity in this model is unusually low among high-redshift galaxies ($\beta$ = 0.7$^{+0.2}_{-0.3}$, cf. 1$-$2.5; \eg \citealt[][and references therein]{Casey12a}). While this directly affects the slope of the Rayleigh-Jeans tail of the modified blackbody function, such low emissivity has insignificant impact on the inferred IR luminosity.
Despite the factor of $\sim$\,2 difference
in dust mass
between models, the derived IR luminosities are comparable. As such, we find no evidence of correlation between the inferred dust mass and IR luminosity.

A comparison of the reduced $\chi^2$ between the two models suggests that the optically thick model fit is superior to the optically thin model fit. In the preferred model, the marginalized PDF of the dust emissivity index is highly non-Gaussian, with a noticeably broad distribution across
the allowed range (upper limit: 2.2). We therefore fit a second optically thick model, changing the upper limit of $\beta$ to 3.0. The difference in each
parameter of the best-fit solution between these two scenarios is less than 3\%, except for the dust mass, which is boosted by a factor of $\sim$\,2. This is
unsurprising given the dearth of data in the rest-frame FIR wavebands, leading to a weakly constrained emissivity, and thus, inferred dust mass.
In our analysis, we employ the best-fit parameters and the corresponding inferred properties (dust mass, and IR luminosity) from the former, where a tighter constraint is imposed on $\beta$.

\end{document}